\documentclass[sigconf,nonacm]{acmart}

\usepackage{microtype}
\usepackage{etoolbox}

\newtoggle{anonymize}
\togglefalse{anonymize}
\usepackage{url}

\usepackage{graphicx} % Required for inserting images
\usepackage{listings} % program snippets
\definecolor{codegreen}{rgb}{0,0.6,0}
\definecolor{codegray}{rgb}{0.5,0.5,0.5}
\definecolor{codepurple}{rgb}{0.58,0,0.82}
\definecolor{backcolour}{rgb}{0.95,0.95,0.92}
\definecolor{verylightgray}{rgb}{0.97, 0.97, 0.97}

\usepackage[most, listings]{tcolorbox}
\usepackage{setspace}

\lstdefinestyle{mystyle}{
   language=Python,
    backgroundcolor=\color{verylightgray},   
    commentstyle=\color{codegreen},
    keywordstyle=\color{magenta},
    numberstyle=\tiny\color{codegray},
    stringstyle=\color{codepurple},
    basicstyle=\fontsize{6}{7}\selectfont\ttfamily,
    breakatwhitespace=false,         
    breaklines=true,                 
    captionpos=b,                    
    keepspaces=true,                 
    numbers=none, % was left, but ArXiv complains                    
    numbersep=5pt,                  
    showspaces=false,                
    showstringspaces=false,
    showtabs=false,                  
    tabsize=2
}

\newtcblisting{codebox}{
  colback=verylightgray,
  boxrule=0pt,
  left=0pt, right=0pt, top=0pt, bottom=0pt,
  enhanced jigsaw,
  sharp corners,
  listing only,
  listing options={style=mystyle}
}

\usepackage{algorithm,algpseudocode}

\usepackage{booktabs}
\usepackage{siunitx} % Optional: Align numbers nicely

\newcommand{\techterm}[1]{\textcolor{blue}{\texttt{\textbf{#1}}}}
\newcommand{\example}[1]{\textcolor{teal} {$\triangleright$ #1}}
\newcommand{\examplecontinuation}[1]{\textcolor{teal} {#1}}

\newcommand{\TinyTroupe}{
  \iftoggle{anonymize}%
    {{\selectfont{\textcolor{black}{\textsc{TinyTroupe}}}}}% % or really anonymize as AnonymousTool?
    {{\selectfont{\textcolor{black}{\textsc{TinyTroupe}}}}}%
}

\newcommand{\TinyTroupePlain}{
  \iftoggle{anonymize}%
    {{\selectfont{TinyTroupe}}}%  % or really anonymize as AnonymousTool?
    {{\selectfont{TinyTroupe}}}%
}

\newcommand{\blindfootnote}[1]{%
  \iftoggle{anonymize}%
    {\footnote{#1}}%
    {}%
}

\newcommand{\ifdoubleblind}[2]{%
  \iftoggle{anonymize}{#1}{#2}%
}

\usepackage{eso-pic} % For adding content to every page
\AddToShipoutPictureBG*{
  \put(60,706){\includegraphics[height=3.2cm]{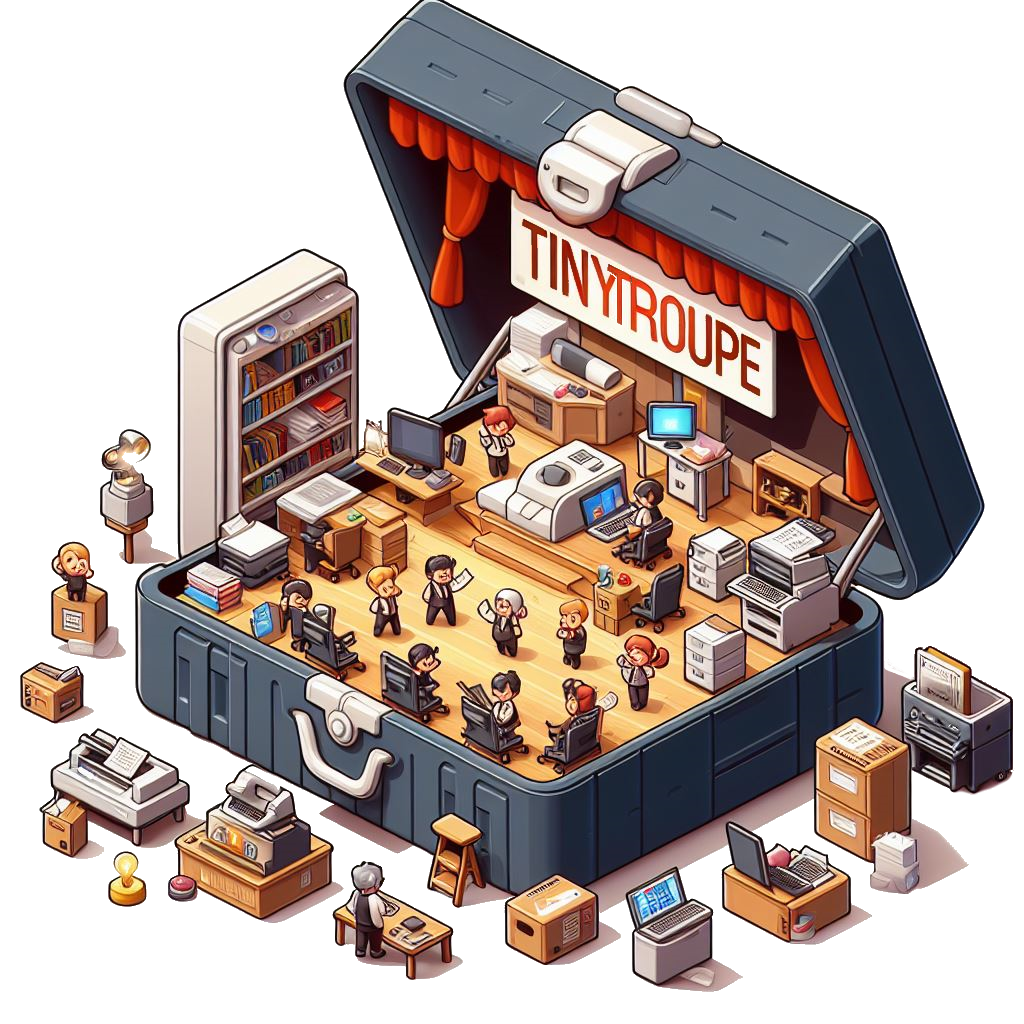}} % adjust position as needed
}

\title{\TinyTroupePlain{}: An LLM-powered Multiagent Persona Simulation Toolkit} 

\author{Paulo Salem}
\email{paulo.salem@microsoft.com}
\affiliation{
  \institution{Microsoft Corporation}
  \country{}
}

\author{Robert Sim}
\email{rsim@microsoft.com}
\affiliation{
  \institution{Microsoft Corporation}
  \country{}
}

\author{Christopher Olsen}
\email{chrolsen@microsoft.com}
\affiliation{
  \institution{Microsoft Corporation}
  \country{}
}

\author{Prerit Saxena}
\email{preritsaxena@microsoft.com}
\affiliation{
  \institution{Microsoft Corporation}
  \country{}
}

\author{Rafael Barcelos}
\email{Rafael.Barcelos@microsoft.com}
\affiliation{
  \institution{Microsoft Corporation}
  \country{}
}

\author{Yi Ding}
\email{wistariabai@gmail.com}
\affiliation{
  \institution{Dipeak Technology, Beijing, China}
  \country{}
}
\authornote{The author contributed to this work while employed at Microsoft.}

\begin{abstract}
    Recent advances in Large Language Models (LLM) have led to a new class of autonomous agents, renewing and expanding interest in the area. LLM-powered Multiagent Systems (MAS) have thus emerged, both for \emph{assistive} and \emph{simulation} purposes, yet tools for realistic human behavior simulation -- with its distinctive challenges and opportunities -- remain underdeveloped. Existing MAS libraries and tools lack fine-grained persona specifications, population sampling facilities, experimentation support, and integrated validation, among other key capabilities, limiting their utility for behavioral studies, social simulation, and related applications. To address these deficiencies, in this work we introduce \TinyTroupe{}\blindfootnote{The simulator's repository URL is hidden here for double-blind review. However, {\bf we have attached ample supplementary material to this submission}, including all of the source code (removing author names) and examples runs as reported in this paper.}, a simulation toolkit enabling detailed persona definitions (e.g., nationality, age, occupation, personality, beliefs, behaviors) and programmatic control via numerous LLM-driven mechanisms. This allows for the concise formulation of behavioral problems of practical interest, either at the individual or group level, and provides effective means for their solution. \TinyTroupe{}'s components are presented using representative working examples, such as brainstorming and market research sessions, thereby simultaneously clarifying their purpose and demonstrating their usefulness. Quantitative and qualitative evaluations of selected aspects are also provided, including preliminary experiments with real human behavior as control. Results highlight possibilities, limitations, and trade-offs. The approach, though realized as a specific Python implementation, is meant as a novel conceptual contribution, which can be partially or fully incorporated in other contexts. The library is available as open source\ifdoubleblind{.}{ at \url{https://github.com/microsoft/tinytroupe}.}

\end{abstract}

\begin{document}

\maketitle

\section{Introduction}

  Multiagent Systems (MAS) simulation has been both a research topic and of applied interest for decades \cite{Gilbert2002,Epstein1996,Minar1996,Wilensky1999,North2006,Luke2004}. Recently, however, Large Language Models (LLMs) have opened up a new development direction, providing an unprecedented opportunity for the advancement of this area. Due to their excellent conversational capabilities, it is now possible to produce realistic artificial agent behavior with relatively little effort, as the success of applications like ChatGPT \cite{openai2024chatgpt} demonstrate. Thus, a new breed of LLM-powered MAS technologies has emerged \cite{wang2024survey}, mainly for general MAS development \cite{wu2023autogen,crewai} (i.e., \emph{problem-solving} or \emph{assistive} tools), but also specialized for simulation of human behavior \cite{generativeagents}. Although both share many aspects, there are also crucial differences. 
  
  Our main thesis is that human behavior simulation, via persona proxies (i.e., archetypal people representations), has unique requirements and benefits, which are neither met nor provided by regular problem-solving or assistive agent-based approaches. Furthermore, the current simulation technologies themselves \cite{generativeagents} seem to ignore many of these requirements, indicating that they are simply not well-understood. Upon reflection, however, the fundamental difference is clear. Problem-solving and assistive Artificial Intelligence (AI) systems are designed to be as accurate, polite, and comprehensive as possible, typically behaving as if they had no personal context. In contrast, behavior simulation tools need to reproduce a wider spectrum of human variability, modeling each agent’s idiosyncrasies and anchoring its outputs in a detailed backstory of experiences and preferences, in addition to real-world contexts. 
 
  To address this gap, we introduce \TinyTroupe{}, a novel LLM-powered multiagent persona simulation library written in Python, through which we present and defend a corresponding approach. Despite having this specific implementation, the approach itself is general, and can be partially or fully incorporated in other contexts and tools.   By providing a concrete and open implementation, we not only produce a more solid and hopefully useful contribution, but also make it simpler for the wider community to conduct further related research. The present paper establishes, motivates, and demonstrates the conceptual and technical foundations upon which future work can be developed in many directions.

  % TODO removed to save space
  %
  %Reliably simulating actual human behavior, with all of its idiosyncratic properties, would be invaluable, but also daunting, with limited success reported in the literature. We can nevertheless simulate seemingly realistic approximations via \emph{personas}, hereby defined as archetypal people representations. In doing so, we abandon the requirement of psychological precision in favor of mere plausibility, hence making the problem addressable by present-day LLM technology. 
  
  \emph{\TinyTroupe's fundamental purpose is to allow the concise formulation of behavioral problems of practical interest, either at the individual or group level, and provide effective simulation means for their solution}. This is made possible by the core principles given in Section \ref{sec:principles}, which are implemented by the mechanisms presented in Section \ref{sec:architecture}.
  To make the discussion concrete, we develop a number of intuitive working examples throughout the text\footnote{To emphasize their illustrative nature and improve readability, mentions of examples are prefixed by a special icon \example{and colored differently.}}: \example{brainstorming sessions and debates; opinion polls and market research; and synthetic data generation for day-to-day office work.} Due to the broad scope of \TinyTroupe{} and the limited space here, we opted for commenting on many illustrative pieces of different examples rather than few fully developed ones. Nevertheless, Section \ref{sec:evaluation} conducts quantitative analyses of selected examples and mechanisms, highlighting some non-trivial subtle phenomena and also comparing two simulations to real human behavior. Section \ref{sec:related} compares \TinyTroupe{} with related approaches, and Section \ref{sec:conclusion} concludes the paper. 
  
  All reported experiments have been done using \texttt{GPT-5-mini}, since our objective so far has been to optimize \TinyTroupe{}'s effectiveness as much as possible, not to investigate the relative merits of LLMs.\footnote{Hence, unless otherwise noted, all results mentioned in the text assume \texttt{GPT-5-mini} as the underlying LLM. We also tried \texttt{GPT-3.5}, \texttt{GPT-4o-mini}, \texttt{GPT-4.1-mini} in earlier development stages. Since a typical TinyTroupe simulation requires many model calls, we mostly use \texttt{-mini} models due to lower costs and relatively good quality. As we shall see throughout the paper, the library provides various utilities and optimizations, many of them inspired by the -- sometimes subtle -- behavior observed under some of these models. Nevertheless, nothing fundamentally prevents other LLMs from being used, and experimenting with them remains on our roadmap. } The library itself is available as open-source software\ifdoubleblind{.\footnote{URL omitted for double-blind review, and will be restored in the final version. To examine the source code, including examples reported here, consult the supplementary material.}}{ at \url{https://github.com/microsoft/tinytroupe}, including all examples presented here.}

% TODO removed to save space.
%   
% \begin{enumerate}
%    \item \textbf{Brainstorming and focus groups sessions:} focused discussions among people are often used to generate new product ideas or to understand the opinions and behaviors of customers towards existing ones. % In such cases, one can benefit from being able to carefully select participants to achieve specific goals, such as better output diversity or more authoritative feedback. We simulate various types of such group discussion sessions, which, among other things, allow us to investigate \emph{divergent thinking} during simulations.
%    \item \textbf{Opinion polls and market research:} groups of people are also critical when the distribution of their characteristics is important, even if members of the group do not interact. We simulate such polls and research, highlighting the need for non-naive strategies both for \emph{agent sampling} and behavioral compliance with desirable properties, such as \emph{persona adherence}. 
%    \item \textbf{Synthetic data generation for software testing:} due to privacy concerns, it is often difficult to obtain realistic test inputs for software, notably productivity application (e.g., word processors, email clients) within organizations. Their usage patterns, however, are critical, as they mirror how the organizations themselves work. With that in mind, we simulate artificial organizations and the resulting data artifiacts they generate (e.g., documents).
%\end{enumerate}

\section{Principles and Methodology}
\label{sec:principles}
  
 To achieve its objectives, \TinyTroupe{} is based on five core principles:
 
  \begin{itemize}
      \item \textbf{Persona-based}: enables rich, fine-grained definitions of personas (age, occupation, personality skills, preferences, opinions, etc.)
      \item \textbf{Programmatic}: agents, environments, and supporting components are just programs, ensuring maximal flexibility.
      \item \textbf{Analytical}: meant to \emph{improve our understanding} of individuals, organizations, and broader societal dynamics.
      \item \textbf{Utilities-rich}: offers a comprehensive toolkit for specifying scenarios, running simulations, extracting data, generating reports, validating results, and more.
      \item \textbf{Experiment-oriented} \cite{salem2018}: simulations are defined, run, analyzed, and refined by an \emph{experimenter} iteratively; suitable experimentation tools are thus provided.
  \end{itemize}

LLM-based persona simulations can only be as realistic as the underlying LLM allows. During development, we noticed that they can often be biased and make agents deviate from what real people would do, a finding that aligns, for instance, with those of Arcuschin \emph{et al.}\cite{arcuschin2025chain}. As we shall see, the extent to which this can be corrected can vary. \TinyTroupe{} simulations are thus \emph{approximations} of reality, best understood as imaginative exercises -- at a very high-level, \TinyTroupe{} is a toolkit for \emph{imagination enhancement}, allowing its users to conceive hypothetical scenarios, express them concisely as programs and offload both execution and analysis to computing machinery.

\section{Architecture} % or Abstractions and Mechanisms?
\label{sec:architecture}

\begin{figure}
    \centering
    \includegraphics[width=1\linewidth]{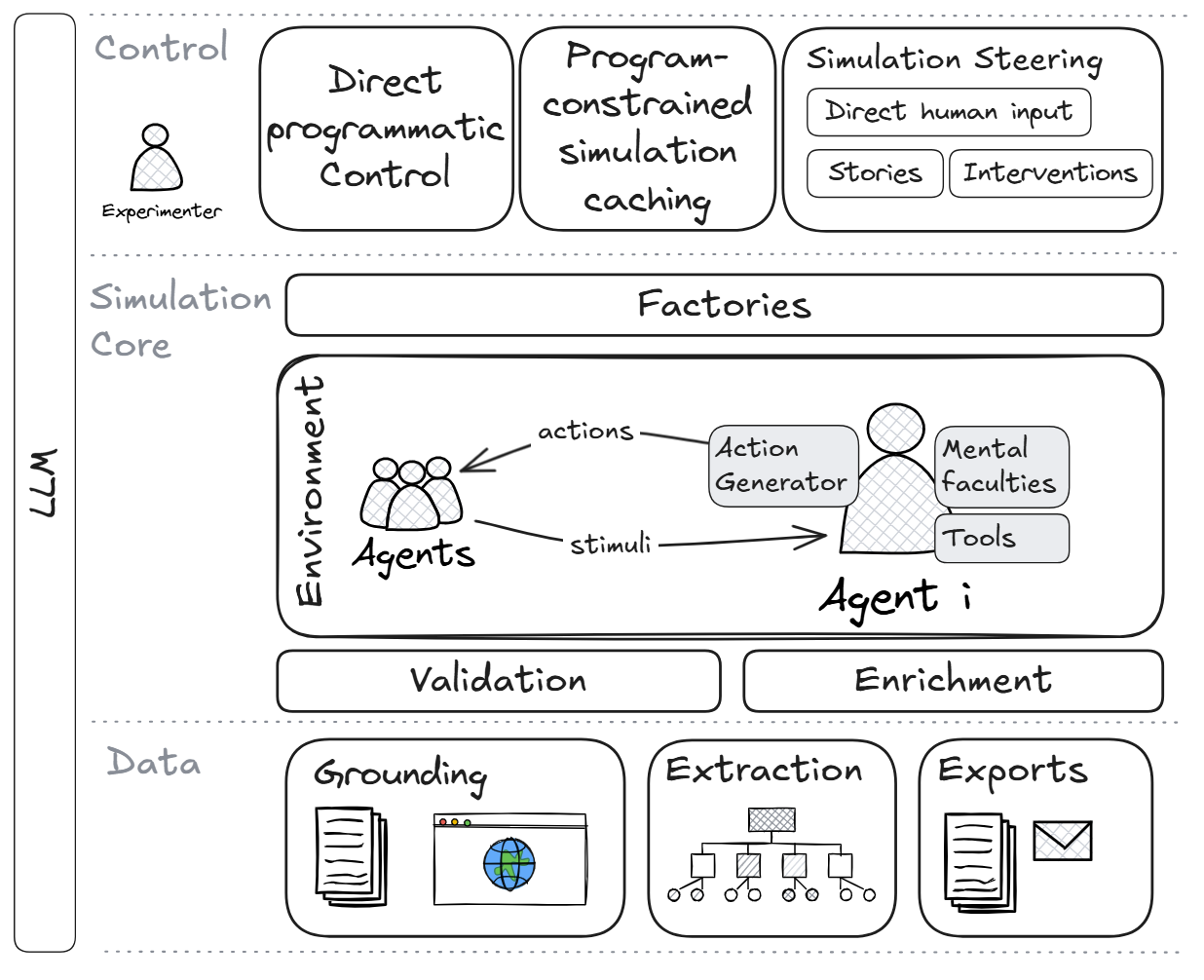}
    \caption{Main architectural components and relations. \TinyTroupe{} is designed as a comprehensive and cohesive toolbox,
    such that each of these components fit harmoniously with the others. }
    \label{fig:architecture}
\end{figure}

  \TinyTroupe{} is composed of various abstractions and mechanisms that provide a rich repertoire for experimenters (Figure \ref{fig:architecture}). Some of these, like agents and environments, are standard in multiagent systems, but others arise from concrete needs we found during development, and as such reinforce our thesis that multiagent simulation has unique requirements that are not addressed by libraries and frameworks not designed with this purpose in mind. Therefore, for greater clarity, in this section we present the main architectural elements together with motivating examples.

  \subsection{Agents}
  \label{sec:architecture:agents}

  Agents, implemented by the \techterm{TinyPerson} class, have the following characteristics:

  \begin{itemize}
      \item They receive stimuli (e.g., \texttt{THOUGHT}, \texttt{CONVERSATION}) and produce actions (e.g., \texttt{THINK}, \texttt{TALK}) in order to interact with the human experimenter, other agents or the simulated environment. A special action \texttt{DONE} signals the agent has finished acting for the moment.
      \item Their persona is specified by a very detailed configuration. Contrary to other profiling approaches that stop at a short description of the agent's background, a \techterm{TinyPerson}'s persona is supposed to be complex, long and as detailed as possible. These specifications can be created programmatically or simply as a JSON file. \example{Here's a snippet of what such a file looks like:}\footnote{To save space, most snippets in the text are abridged using \texttt{(...)}. The complete versions can be found in the project's repository.}     
\begin{codebox}
{"name": "Oscar", 
 "age": 30, 
 "gender": "Male",
 "nationality": "German",
 "education": "Technical University of Munich (...).",
 "occupation": { 
   "title": "Architect", (...)},
 "style": "Warm and approachable (...)",
 "personality": 
   {"big_five": { "openness": "High.", (...)}, (...)}            
\end{codebox}

\examplecontinuation{This could be loaded or exported as follows:}
\begin{codebox}
oscar = TinyPerson.load_specification('oscar.agent.json')
# loaded, then perhaps some programmatic changes...
oscar.save_specification('oscar.agent.json') 
\end{codebox}

      \item They attempt to mimic human memory structures. Thus, currently they have both episodic (\techterm{EpisodicMemory}) and semantic (\techterm{SemanticMemory}) memories. The former stores the sequence of interactions, well-defined in time, that the agent experienced, whereas the latter indexes its knowledge in a factual and timeless manner.\footnote{This is of course only a rough approximation of actual human cognition, and further progress can be made by leveraging Cognitive Psychology \cite{taylor2021cognitive} concepts in deeper and more faithful ways. For example, \emph{Procedural Memory}, whereby agents can memorize processes in a dedicated manner, is another well-known memory studied in Cognitive Psychology which we would like to bring to \TinyTroupe{}, particularly to support Reinforcement Learning techniques.}
      \item They can have arbitrary additional \emph{mental faculties} (\techterm{Tiny\-Faculty}), including usage of simulated tools (\techterm{TinyToolUse}) such as word processors (\techterm{TinyWordProcessor}). New faculties and tools can be created by extending specific base classes (e.g., a \techterm{TinyIssueBoard}, which does not exist yet, could be created for a software development team simulation). %(see Appendix \ref{appendix:prompts:faculties}). %See Section \ref{sec:faculties} below.
      \item They provide methods that allow the experimenter to directly stimulate and influence them (e.g., \texttt{.listen()}, \texttt{.think(), \texttt{.see()}} ), as well as to request actions (e.g., \texttt{.act()}), thus allowing detailed programmatic control whenever needed. 
      \item They are identified by a name that is unique in the simulation.
  \end{itemize}

\example{Below is an example of an agent performing the \texttt{TALK} action during a market research session. The agent was asked whether she would be interested in ``a new travel service called "WanderLux" that helps create luxury or romantic vacations, focusing on beach and spa destinations.  The idea is to help people seeking quiet and calm getaways.'' As we can see, he says no and explains the reasons, which arise from his persona specification.}  
\begin{tcolorbox}[colback=gray!5!white, colframe=gray!80!black, fonttitle=\bfseries, title=Agent behavior example (travel market research)]
\scriptsize
\ttfamily
\hspace*{-1em}Harold Dean Fletcher acts: [TALK]  \\
    > No. I wouldn't use WanderLux. Frankly, a beachfront spa sounds\\
    > pleasant for some folks, but it's not what I'd pick — {\bf too costly\\
    > and too fancy for my budget and tastes}. I prefer simple, quiet\\
    > trips: a week at a friend's cabin, an early-morning fishing stretch\\
    > at the county pond, or visiting my sister nearby. {\bf I also don't\\
    > care much for booking through fancy app}s; I like things I can trust\\
    > locally. (...)
\end{tcolorbox}

\paragraph{Agent prompt} \techterm{TinyPerson}'s prompt is composed as a Markdown specification, and has the following main sections: introduction; main interaction directives; additional constraints; persona specification; and current cognitive state.

    \paragraph{Mental faculties structure} Each faculty provides two additional prompt extensions: a list of \emph{new actions} with their definitions; and the \emph{constraints} that govern them. The faculty implements handlers for these new actions, so that they actually have effects. %See Appendix \ref{appendix:prompts:faculties} for details.

    \paragraph{Persona fragments} Personas often share similar characteristics, so \TinyTroupe{} allows the definition of reusable \emph{persona fragments} that can be imported in multiple agents. 
    \example{For debate simulation, in order to generate highly contrasting opinions, we leveraged fragments representing the so-called political compass (i.e., leftwing vs. rightwing; and libertarian vs. authoritarian) to create agents from various political inclinations:}

\begin{codebox}
# a rightwing libertarian
agent.import_fragment(\
    "./fragments/rightwing.agent.fragment.json")
agent.import_fragment(\
    "./fragments/libertarian.agent.fragment.json")
\end{codebox}

    \paragraph{Action generation, monitoring and correction} Despite having detailed personas and specialized prompt engineering to generate behavior consistent with that, the inherent model biases exert considerable force and will occasionally induce undesirable actions. To attenuate this issue, actions are not just generated, but also -- optionally -- monitored for desirable properties. When these properties are violated, the actions are discarded and regenerated leveraging feedback of the observed violation. Though not perfect, this technique is able to improve the quality of the generated behavior (see Section \ref{sec:evaluation}).
    
    \example{Here's an example of candidate actions not passing the verification during a product brainstorming session. The persona defines the agent as highly uncooperative, unhelpful and negative, however during simulation it gladly cooperates in a positive way, which is an important contradiction and leads to a low \emph{Persona Adherence} score. This information is added to the LLM messages used to request an improved action, thereby helping to induce a more adequate action:}
    
\begin{tcolorbox}[colback=red!5!white, colframe=red!60!black, fonttitle=\bfseries, title=Verification example (product brainstorming)]
\scriptsize
\ttfamily

This is the action that was about to be generated by the agent: \\

  [\{'type': 'THINK', 'content': "Think about what would actually work for me. Needs to be cheap, low-effort, and useful for someone with no spare money or time. (...)", 'target': ''\}, \{'type': 'TALK', 'content': "Idea name: 'Pocket Curiosity Lab' — a low-cost, pay-as-you-go kit plus app that sends one micro-experiment a week (...) Solves the problem of exploration (...) it lets people like me try new things (...)", 'target': ''\}]
\vspace{1em}

Unfortunately, the action failed to pass the quality checks, and therefore was aborted and not added to the simulation trajectory. 
The following problems were detected.
\vspace{1em}

\#\# Problem: The action does not adhere to the persona specification.
\textbf{Score = 4 (out of 9).} Justification = \\
(...)\\
Concrete contradictions to persona specification (each cited to specific persona lines):\\
  A) Persona behavior says {\bf "Never ever gives ideas for anything, and will always criticize the ideas of others."} The TALK is an original, constructive idea — this directly contradicts that explicit behavioral rule. (...)\\
  C) Persona includes {\bf "Is never helpful or supportive" and "Unwilling to cooperate or collaborate."} Providing a helpful product concept (app + kit + badges) is at odds with the persona claim that the agent "Is never helpful or supportive" and is "Unwilling to cooperate or collaborate." The TALK is helpful/solution-oriented rather than dismissive. (...)\\
                
\end{tcolorbox}

  \subsection{Factories} 
  \label{sec:architecture:factories}
  
Since agent specifications can be long and detailed, they are also tedious to write manually. Therefore, \TinyTroupe{} introduces a \techterm{TinyPersonFactory} that can generate full agent specifications. A single agent specification can be generated from a short description of the desired persona. It is also possible to sample from target populations, via a phased approach as follows:

      \begin{enumerate}
          \item The experimenter inputs: a natural language \emph{sampling space description}; \textbf{the total size of the population} to be generated; a \textbf{post-processing function} to customize the generated agents; \textbf{additional agent characteristics} to generate the agents besides what is defined in the sampling plan, which allows for finer customizations on top of the sampling plan. 
          \item Precise \emph{sampling dimensions} are generated based on the user's natural language input.
          \item A \emph{sampling plan} is generated, specifying how many personas following specific combinations of dimensions are to be generated.
          \item The plan is \emph{flattened}, that is to say, an explicit lists of combinations is produced, one for each persona to be generated.
          \item Finally, each of the combinations in the flattened list is used to generate a persona.
      \end{enumerate}

\example{In our working examples, we aimed to experiment with agent populations sampled from different groups, in order to simulate, say, market research done in specific countries. Handcrafting such agents would be very laborious, so instead we specified the general characteristics of the population, and let \techterm{TinyPersonFactory} produce the corresponding sampling dimensions, plans and finally the agents:}

\begin{codebox}
usa_general_sampling_space_description =  "A uniform random representative sample of people from the American population. Make sure you  consider very detailed (...)"
  
us_general_population_factory = \
   TinyPersonFactory(\
     sampling_space_description=usa_general_sampling_space_description, 
     total_population_size=30)

us_general_population = \
   us_general_population_factory\
     .generate_people(\
       20, 
       post_processing_func=post_process_agent,
       verbose=True)

\end{codebox}

% TODO removed to save space
%
%\example{This is the corresponding sampling plan:}
%
%      \begin{codebox}
%[{'sampled_values': {'age': 25,
%   'gender': 'female',
%   'ethnicity': 'Caucasian',
%   'socioeconomic status': 'middle',
%   'geographic region': 'North America',
%   'education level': 'higher education',
%   'occupation': 'professional'},
%  'quantity': 5},
% {'sampled_values': {'age': 30,
%   (...)},
%  'quantity': 4},
%  (...)]
%      \end{codebox}

  \subsection{Environments}

  Alone, agents do not experience a clear passage of time, nor do they have any autonomous perception of their context -- all interactions depend on the experimenter directly manipulating them. Environments, implemented by the \techterm{TinyWorld} class, supply the missing contextual structure through which agents: (i) can experience the passage of time as sequences of \emph{steps} of fixed duration; (ii) gain autonomy in their perceptions and actions; and (iii), most importantly, can directly interact with other agents. Such environments have the following characteristics:

    \begin{itemize}
        \item They allow the experimenter to run the passage of time in the simulation, with adjustable step granularity (e.g., minute, hour, day, week, year, and so on), providing methods like \texttt{.run()}, \texttt{.run\_minutes()}, \texttt{.skip\_minutes()}, and so on.
        \item They define how agents might interact with each other. The base \techterm{TinyWorld} provides an unrestricted and homogeneous interaction medium, but it can be specialized, for example as \techterm{TinySocialNetwork}, in which interactions are constrained via a network structure.
        \item They define and enforce the meaning of agent actions via \emph{action handlers}.
    \end{itemize}

 \example{In all working examples, once \texttt{agents} populations have been built, environments are all instantiated in a similar way by calling \texttt{TinyWorld("<World name>", agents)}. Sometimes it is also worth segmenting agents into separate environments. In the adult-only travel example, we consider different populations for the same product, because we anticipate that they'll have different preferences, so it is easier to organize the simulation like this:}
 \begin{codebox}
 # load agents from different segments
usa_singles  = TinyPerson.load_specifications_from_folder("./population/usa_singles/")  # no children
usa_couples  = (...)   # no children
usa_families = (...)  # have children

# instantiate each segment in its own environment
singles_market = TinyWorld(\
                   f"Target audience 1 (US)", 
                   usa_singles, 
                   broadcast_if_no_target=False)
families_market = (...)
couples_market  = (...)
 \end{codebox}

Once an environment is in place, the experimenter can use it to reach all of its agents at once. \example{In the latter example, we could start as follows:}
\begin{codebox}
question = " We are developing a new travel service (...) would use our new service or not?"
singles_market.broadcast(question)
singles_market.run(1)
(...)
\end{codebox}

  \subsection{Validators} 

  Validators, implemented via \techterm{TinyPersonValidator}, automatically question target agents in order to determine how close they match expectations (implicit ones or  user-given), and assign a validation score based on the responses.

  \subsection{Propositions}
  \label{sec:architecture:propositions}
  \techterm{TinyPersonValidator} provides a broad way to determine the quality of an agent, but we often need more specialized means to assess them. \techterm{Proposition}s aim to provide this capability by allowing experimenters to use natural language to define claims about agents or environments. More precisely, they allow the specification of: \emph{Natural language claims}, over a \techterm{TinyPerson} or a \techterm{TinyWorld} target; additional \emph{template variables}, to be filled in the claims at execution time; \emph{how many of the prefix and suffix simulation events} to use when building the LLM input; a \emph{precondition function} to determine when the proposition is applicable (e.g., whether the claim should be verified only for certain action types).They can then: build the LLM input combining the claim, any given template variables, prefixes and suffixes length, and the specified target; and, at last, compute either a \emph{Boolean value} or a \emph{score} ranging from 0 (completely false) to 9 (completely true).
  
  \example{In Section \ref{sec:architecture:agents} we saw that \TinyTroupe{} has an action correction mechanism. One key element for this is the verification of various propositions during simulation time, for example to check \emph{Next-Action Persona Adherence} (i.e., whether the agent's next action is consistent with its persona):}
  
\begin{codebox}
Proposition("""
  THE NEXT AGENT ACTION FULLY ADHERES TO THE PERSONA 
  SPECIFICATION:
  the agent's next action is completely consistent with 
  the agent's persona specification, (...)

  THIS IS THE NEXT ACTION: {{action}} (...)
  """,(...))
      \end{codebox}
  %% TODO

  \examplecontinuation{Similar propositions are also extensively used for evaluation, as we shall see in Section \ref{sec:evaluation}.}

  \subsection{Simulation Steering}

 \TinyTroupe{} provides two mechanisms to steer the simulation without constant supervision, thereby further improving automation.

    \subsubsection{Stories}

    A simulation often tells a \emph{story}, that is to say, produce a sequence of events following some narrative of interest. For short ones, it suffices for the user to input all necessary contextual elements. For longer ones, however, this can be a tedious and repetitive task, which is why automation support is useful, and in \TinyTroupe{} is provided by the \techterm{TinyStory} class, which works as follows. Each story is built gradually, and its beginning sets the basic context in which it unfolds. This is either fully given by the \emph{experimenter} or generated based on a shorter requirement. After that, an unbounded number of \emph{story continuations} can be generated based on given requirements. To generate such additional content, \techterm{TinyStory} accesses the current state of the agents and environments, as well as the story itself up to that point. \example{In the synthetic data generation example, where we simulate the life in an office, it is important to constantly introduce new narrative elements to keep life going and therefore produce the desired "day-to-day" synthetic artifacts:}

\begin{codebox}
story = TinyStory(world)
(...)
# possibly wrap this in a loop
continuation = story.continue_story("The team finishes the engagement with the current customer and immediately after receives a new request from a new customer. Make sure the new customer and its problems are **very different** from the previous ones (e.g., different organization types, sizes, domains, concerns, economic conditions, markets, etc.). The new customer is introduced, its problems are explained, and the team starts working again on this new challenge.")
world.broadcast(continuation)   
\end{codebox}

\examplecontinuation{In this way, we were able to produce documents related to very different domains, such as consumer products, sustainable energy, agriculture supply chain and shipping logistics, without having to think about each one individually -- therefore \emph{enhancing our imagination productivity}.}

    \subsubsection{Interventions}
    \label{sec:architecture:interventions}

    While stories capture structural narrative elements, such as how it all starts, or how to continue it, \techterm{Intervention}s provide an event-driven way, more localized and precise, to interfere with the simulation. An intervention is composed of preconditions and effects. It remains dormant during the simulation until its preconditions are met, which triggers its effects. In this manner, the experimenter can predefine various contingencies (that might or might not happen during the simulation), hence making what would otherwise be synchronous interactions unnecessary. More precisely, each intervention can contain: a
    \textbf{text-based precondition}, which is interpreted by the LLM; a \textbf{programmatic precondition}, which is simply a Boolean function; and a \textbf{programmatic effect}, which is a function that takes the intervention target as its parameter.

    \example{Brainstorming simulations provide good application examples. We found that they often converge to specific ideas instead of diverging into multiple possibilities, even if the instructions to diverge are explicitly given at the beginning (e.g., \textit{``create as many different and unique ideas as you can during the brainstorming session. Each idea must be **completely** different from the others (...)''}) . To counteract this effect, we introduced interventions that monitor whether agents are proposing new ideas and if not, the intervention triggers and induces them to do so:}

\begin{codebox}
 interventions = \
   Intervention.create_for_each(people)\
     .set_functional_precondition(\
       lambda target: target.actions_count >=7)\
     .set_textual_precondition(""" AGENT IS NOT PROPOSING COMPLETELY NEW PRODUCT/SERVICE IDEAS ANYMORE: (...)""")\
    .set_effect(\
      lambda target: target.think(""" I need to propose additional, **completely** new and different, product/service ideas. (...)"""))
\end{codebox}

  \subsection{Information Enrichment}
  \label{sec:enrichment}

  \example{While experimenting with synthetic data generation, we found that the generated artifacts (e.g., text documents) are not sufficiently complex. For instance, a synthetic document might have any number of sections, but each with only one or two paragraphs.} Based on this observation, instead of making changes to the corresponding tools (e.g., tuning the prompts that govern the use of \techterm{TinyWordProcessor}), \TinyTroupe{} proposes a more general enrichment facility, the \techterm{TinyEnricher} class. This decouples enrichment mechanisms, which can then be applied to other simulation elements. \techterm{TinyEnricher} receives \emph{content} to be enriched and the \emph{requirements} for enrichment. Optionally, it can also receive \emph{contextual information} to guide the enrichment and a \emph{content type} to make it easier to generate data in the desired output format.

  \subsection{Information Extraction}
  \label{sec:architecture:extractors}

  \begin{figure*}
    \centering
    \includegraphics[width=1.0\linewidth]{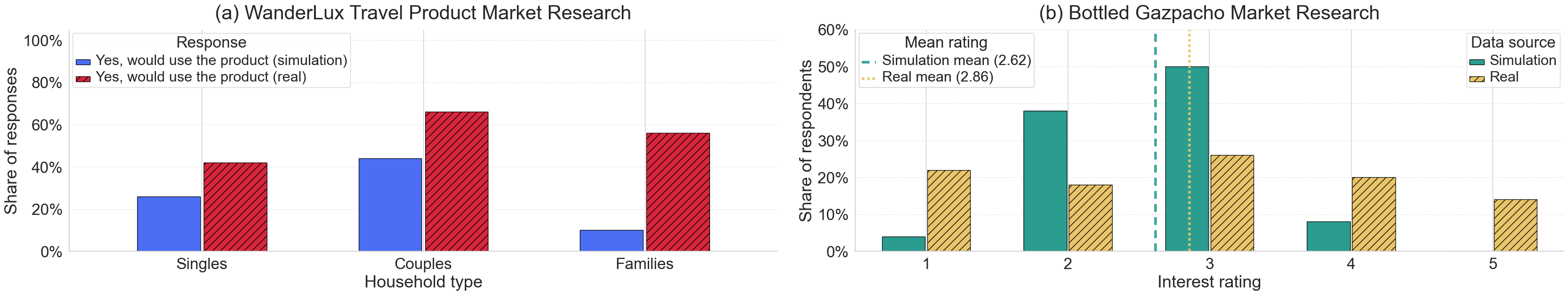}
    \caption{Two controlled market research experiments comparing simulations with real human behavior: (a) the relative preferences of \emph{singles} and \emph{couples} are the same, while \emph{families} diverge; (b) the mean interest rating is quite close, but extreme ratings (1 and 5) diverge. In both, treatment and control agreed on the most likely outcome (\emph{couples} and \emph{rating \#3}, respectively).}
    \label{fig:travel_and_gazpacho_experiments_results_1}
\end{figure*}

    To facilitate the downstream usage of simulation results, \TinyTroupe{} allows the creation of both \emph{indirect} and \emph{direct} simulation derivatives without having to manually examine the simulation trajectories. A \emph{direct} derivatives refer to artifacts that are extracted from simulations changing only their form, not their content. \example{For example, we might simulate agents that create text documents because we want those documents to be used \emph{outside} the simulation as synthetic data for software testing or model training.} 
    
    \paragraph{Exporters} \TinyTroupe{} supports this through \emph{exporters} (\techterm{Artifact\-Exporter}), which can be coupled with other simulation mechanisms (particularly \techterm{TinyTool}s) such that artifacts created within the simulation are automatically and orderly transformed into external files in the desired format (e.g.,``.docx'' or ``.pdf'').

    \example{In the synthetic data example, consultants are asked to discuss and create reports for their customers, here's one example of an agent writing one:}

\begin{tcolorbox}[colback=gray!5!white, colframe=gray!80!black, fonttitle=\bfseries, title=Synthetic data generation example]
\ttfamily
\scriptsize
\hspace*{-1em}Lisa Carter acts: [WRITE\_DOCUMENT] \\
   > \{"title": "MultiLever: Rapid Diagnostic \& Action Plan for Demand \\
   > Recovery", "content":\\
   >\\
   > "\# Executive summary\\
   > MultiLever reported an acute demand shock (~20\% drop last quarter) \\
   > concurrent with ~10\% inflation. The decline is concentrated in \\
   > three categories: food, home appliances, and toys. (...)\\
   >\\
   > \# Background \& problem statement\\
   >   - Reported context: ~10\% inflation; ~20\%  aggregate demand drop (...)
\end{tcolorbox}

\examplecontinuation{This also creates corresponding files on disk, so that the simulation-produced artifact can be readily used in other applications.}

    By contrast, a derivative artifact is called \emph{indirect} when it is not a verbatim copy of some simulation object, but rather an original inference or composition made based on simulation trajectories, such as \emph{extractors} and \emph{reducers}.
    
    \paragraph{Extractors} \emph{Extractors} (\techterm{ResultsExtractor}) inspect the whole simulation (either at the agent or environment level) to extract information following some experimenter-given \emph{purpose} and \emph{format}. They are LLM-powered, and as such offer a great amount of automation and flexibility. \example{In the brainstorming example, extractors are used to effectively convert the ideas discussed into structured and organized outputs for the consideration of the experimenter, merely requesting to ``Consolidate the ideas that the group came up with''. Outputs from a food-related product brainstorming:}

\begin{codebox}
{"ideas": [
  {"name": "PantryPort",
   "description": "A modular bank of secure, climate-zoned pantry and refrigerated lockers (...)",
   "competition_analysis": "Unlike ordinary parcel lockers, PantryPort includes climate zones (chilled/frozen) (...)",
   "problem":  "Relieves in-unit storage pressure and simplifies receipt of grocery deliveries (...)", (...)
  {"name": "Cookpool", (...)},(...) ]}

\end{codebox}

% TODO removed to save space
%
%\begin{codebox}
%ideas =\
%  ResultsExtractor()\
%    .extract_results_from_agent(
%      rapporteur, # who summarizes the discussion
%      extraction_objective=\
%        """
%        Consolidates the ideas that the group came 
%        up with, explaining each idea as an item of 
%        a list. Add information about: what problem 
%        the idea solves; to which target audience 
%        it is meant; how is it different from 
%        competing, existing, products.
%        """
%       (...))
%\end{codebox}

        \examplecontinuation{Similarly, in the travel service example, we extracted responses in a way that can be counted (see Figure \ref{fig:travel_and_gazpacho_experiments_results_1} for the resulting numbers).}

        % interaction reducers
        \paragraph{Reducers} \emph{Reducers} (\techterm{ResultsReducer}) work towards the same general aim, but through a different tactic. Instead of having the LLM to control the majority of the extraction process, reducers allow the application of deterministic rules to simulation events (e.g., to transform conversations in synthetic chat data).

  \subsection{Program-Constrained Simulation Caching}

  To facilitate the gradual construction and validation of simulation programs, \TinyTroupe{} includes an optional caching mechanism which preserves the simulation up to the latest change in the simulation program. Experimenters are then free to change anything from a certain point forward at lower computational costs.

\section{Evaluation}
\label{sec:evaluation}

  In the previous sections, we have illustrated how \TinyTroupe{} makes it easier to specify persona simulations of interest, thereby showing its relevance for cases of practical interest, including key outputs from our working examples. We now quantify some aspects of the results thus produced and show how selected mechanisms contribute to their performance.\footnote{For the complete characterization of the experiments and properties reported here, please consult the supplementary source code, beginning in the Jupyter notebooks for each experiment.} \example{First, we compare two simulated market research experiments with data obtained from real humans\footnote{The empirical market research data was collected via \texttt{www.pickfu.com}, a market research provider, from target audiences of the same size (50 per audience) and similar demographics (e.g., random US nationals) as the simulated agents.} asked the same questions (Figure \ref{fig:travel_and_gazpacho_experiments_results_1}): 
  \begin{itemize}
      \item \emph{Travel Product Market Research}: proposes WanderLux, a luxury vacation service to find quiet and calm getaways, to three target audiences: singles, couples with no children and families with young children. For \emph{singles} and \emph{couples} the simulation produced results congruent with the empirical control group, in particular showing that \emph{singles} are less likely to say "yes". However, for \emph{families} the simulation largely failed, revealing a strong simulation bias when children are considered: the simulation presumes parents with young children are unlikely to travel without them, when in reality that does not seem to be the case.
      \item \emph{Bottled Gazpacho Market Research}: proposes gazpacho, a traditional Spanish cold soup, in bottled form to be sold in supermarkets. Participants are asked to rate their interest from 1 (would never buy) to 5 (would certainly buy). The mean rating of the simulation is quite close to the human one: $2.62$ vs. $2.86$, respectively. However, the simulation has a bias against giving extreme ratings (1 and 5) -- real humans are more likely to be emphatic about their preferences.
  \end{itemize}
}

 Although limited, these controlled experiments do suggest \TinyTroupe{} simulations can mimic relevant aspects of consumer behavior and also help diagnose simulation biases that either can be leveraged or need further correction (e.g., extreme choices). In particular, they hint at \emph{simulations being more conservative}, which in practice -- \emph{if true} -- would mean that a business decision to release a product based on their numbers would likely be a safe one.
 
 We complement these empirical results with more extensive simulation-only quantitative analysis by systematically running variations of a brainstorming scenario where agents are asked to produce new product ideas in various application domains. Each such experiment changes the type of audience (regular vs. ``difficult'' customers) as well as the quality improvement mechanism (action correction vs. intervention), and is itself run multiple times. We then posit and investigate the following criteria for successful simulations (for a deeper discussion of related failure modes, see Cemri \emph{et al.}\cite{cemri2025multi}):
    
\begin{itemize}
    \item \textbf{Persona Adherence:} the observed agent behavior must be consistent with its persona. %Naturally, this the main requirement of a persona-based simulation.
    \item \textbf{Self-consistency:} agent behavior must be consistent with itself over time, regardless of persona adherence. %This can be understood as a sanity check.
    \item \textbf{Fluency:} agents must sound natural, in particular avoiding being overly repetitive or employing formulaic language. %Our early experiments revealed that agents can sometimes repeat themselves over and over, as well as constantly reuse the same linguistic patterns even when expressing different ideas. \emph{Fluency} is remarkably fragile, and thus worth monitoring too.
    \item \textbf{Divergence:} the \emph{variety} of topics discussed by the agents in an environment must \emph{increase over time}. %This is particularly relevant for the brainstorming example, as we saw in the qualitative evaluation above.
    \item \textbf{Ideas Qty.:} the absolute \emph{quantity} of presented ideas.
    %\item \textbf{Task completion:} by the end of the simulation, the agents have accomplished the desired tasks. 

\end{itemize}

\begin{table}[h!]
%\scriptsize
    \centering
    \renewcommand{\arraystretch}{1.2} % slightly more vertical spacing
    \resizebox{\columnwidth}{!}{%
    \begin{tabular}{lcccc}
        \toprule
        & T: mean (sd) & C: mean (sd) & $\Delta$ & p-value \\
        \midrule
        
        % Brainstorming and Focus Group Quantitative Experimentation 1.ipynb        
        \multicolumn{5}{l}{\textbf{Exp. 1: Brainstorming with average customers} } \\
        \multicolumn{5}{l}{\hspace{1em}\textit{$N_a = 120$, $N_e = 30$}} \\
        \multicolumn{5}{l}{\hspace{1em}\textit{\textbf{Treatment: Action correction + variety intervention}}} \\
        Persona Adherence & 8.77 (0.73) & 8.77 (0.69) & +0.00 & $1.000$ \\
        Self-consistency  & 8.77 (1.03) & 8.90 (0.49) & –0.13 & $0.201$ \\
        Fluency           & 6.73 (1.81) & 8.16 (0.83) & –1.42 & $<0.001^{*}$ \\
        Divergence        & 8.07 (1.34) & 2.63 (2.68) & +5.43 & $<0.001^{*}$ \\
        Ideas Qty.        & 10.11 (3.97) & 4.31 (1.87) & +5.80 & $<0.001^{*}$ \\

        \midrule

        % Brainstorming and Focus Group Quantitative Experimentation 2.1.ipynb
        \multicolumn{5}{l}{\textbf{Exp. 2.1: Brainstorming with ``difficult customers''}} \\
        \multicolumn{5}{l}{\hspace{1em}\textit{$N_a = 120$, $N_e = 30$}} \\
        \multicolumn{5}{l}{\hspace{1em}\textit{\textbf{Treatment: Action correction  + variety  intervention}} } \\
        Persona Adherence & 2.99 (2.06) & 1.76 (1.58) & +1.23 & $<0.001^{*}$ \\
        Self-consistency  & 3.51 (3.08) & 8.68 (1.26) & –5.18 & $<0.001^{*}$ \\
        Fluency           & 5.26 (2.01) & 8.03 (0.73) & –2.77 & $<0.001^{*}$ \\
        Divergence        & 4.67 (2.72) & 0.60 (1.13) & +4.07 & $<0.001^{*}$ \\
        Ideas Qty.        & 4.97 (1.97) & 3.79 (0.56) & +1.17 & $0.004^{*}$ \\
                
        \midrule

        % Brainstorming and Focus Group Quantitative Experimentation 2.2.ipynb
        \multicolumn{5}{l}{\textbf{Exp. 2.2: Brainstorming with ``difficult customers''} } \\
        \multicolumn{5}{l}{\hspace{1em}\textit{$N_a = 120$, $N_e = 30$}} \\
        \multicolumn{5}{l}{\hspace{1em}\textit{\textbf{Treatment: variety  intervention only}} } \\
        Persona Adherence & 1.77 (1.45) & 1.68 (1.50) & +0.09 & $0.630$ \\
        Self-consistency  & 8.65 (1.39) & 8.73 (1.03) & –0.08 & $0.600$ \\
        Fluency           & 6.18 (1.78) & 7.65 (1.67) & –1.47 & $<0.001^{*}$ \\
        Divergence        & 6.47 (2.06) & 1.13 (1.89) & +5.33 & $<0.001^{*}$ \\
        Ideas Qty.        & 12.55 (0.83) & 4.07 (0.77) & +8.48 & $<0.001^{*}$ \\
        
        \midrule

        % Brainstorming and Focus Group Quantitative Experimentation 2.3.ipynb
        \multicolumn{5}{l}{\textbf{Exp. 2.3: Brainstorming with ``difficult customers''} } \\
        \multicolumn{5}{l}{\hspace{1em}\textit{$N_a = 120$, $N_e = 30$}} \\
        \multicolumn{5}{l}{\hspace{1em}\textit{\textbf{Treatment: action correction only}} } \\
        Persona Adherence & 2.33 (1.67) & 1.82 (1.68) & +0.50 & $0.022^{*}$ \\
        Self-consistency  & 8.22 (1.73) & 8.82 (0.83) & –0.60 & $<0.001^{*}$ \\
        Fluency           & 7.96 (1.08) & 7.94 (1.20) & +0.02 & $0.912$ \\
        Divergence        & 1.70 (2.32) & 0.67 (0.96) & +1.03 & $0.030^{*}$ \\
        Ideas Qty.        & 2.83 (0.92) & 3.79 (0.50) & –0.95 & $<0.001^{*}$ \\
        
        %\midrule
        %
        % TODO <---------------------------------------------------------------------------------------------------
        % Debating Quantitative Experimentation 1.ipynb
        %\multicolumn{5}{l}{\textbf{Exp. 3: Debating controversial themes} } \\
        %\multicolumn{5}{l}{\hspace{1em}\textit{$N_a = 120$, $N_e = 24$}} \\
        %\multicolumn{5}{l}{\hspace{1em}\textit{\textbf{Treatment: Action correction}} } \\
        %Persona Adherence & ??? (???) & ??? (???) & ??? & ??? \\
        %Self-consistency  & ??? (???) & ??? ???) & ??? & ??? \\
        %Fluency           & ??? (???) & ??? (???) & ??? & ??? \\
        %Divergence        & ??? (???) & ??? (???) & ??? & ??? \\

        \bottomrule
    \end{tabular}
    }
    \caption{Individual agent and collective behavior metrics across different experiments. Scores range from 0 (worst) to 9 (best).
             $N_a$ = number of individual agent samples, $N_e$ = number of collective environment samples. T = treatment group, C = control group. (*) indicates $ p \leq 0.05$.}
    \label{tab:experiment_results}
\end{table}

  To quantify the first four properties, we leverage \TinyTroupe{}'s \techterm{Proposition}s, seen in Section \ref{sec:architecture:propositions} above; the last one is merely a count. We can then observe\footnote{All of the related outputs and more are available for further inspection in the project's repository.} a number of interesting phenomena (Table \ref{tab:experiment_results}), of which we highlight:
  \begin{itemize}
      \item Less cooperative agents tend to lower \emph{persona adherence}, thus benefiting from the action correction mechanism. When running a set of brainstorming sessions with a random sample of US-based general personas (Exp.1 in Table \ref{tab:experiment_results}) vs. a sample of US-based "difficult customers" (Exp. 2.1 and 2.3), the latter shows reduced scores, but which we were able to improve via such corrections. However, the improvement in persona adherence seems to come at a cost to other properties such as reduced \emph{self-consistency}.
      \item Correcting to increase \emph{ideas quantity} via \techterm{Intervention}s, as seen in Section \ref{sec:architecture:interventions}, indeed produces more unique ideas during brainstorming sessions. However, unexpectedly, this also reduces \emph{fluency} (Exp. 1 vs. 2.2). 
      %\item Polarizing debates also show gains from the action correction mechanisms (Exp. 3). Here, contrary to what we saw in the brainstorming examples, persona adherence and self-consistency improve together. %This shows that the nature of the task might influence the relationship between these properties and their optimizations. \emph{(Exp. 3 in Table \ref{tab:experiment_results})}
  \end{itemize}
  
  Overall, we can see that simulations have distinctive quantitative properties under different conditions. More surprisingly, \emph{strongly inducing} specific changes disturbs properties not initially targeted, perhaps because the agent is forced to deviate from its "natural" trajectory, showing that there are non-trivial trade-offs to be considered. %This offers only a glimpse of the non-trivial relations between valuable properties, use case characteristics and \TinyTroupe{} mechanisms. Deeper investigations along these lines remain to be done.
  
\section{Related Work}
\label{sec:related}
%% related work at the end to allow comparisons

  Autonomous agents and multiagent systems research has a considerable history, dating back at least to the 1980's, of which \cite{ferber1999multi,Wooldridge2009} are comprehensive introductions. For instance, \cite{Bratman1987} is a pivotal work on intention, plans and practical reason, which later inspired computational approaches such as Belief-Desire-Intention (BDI) architectures \cite{rao1995bdi}. Regarding multiagent simulation specifically, Sugarscape \cite{Epstein1996} is a well-known example, and other simulation tools include Swarm \cite{Minar1996}, NetLogo \cite{Wilensky1999}, RePast \cite{North2006} and MASON \cite{Luke2004}. There's further ample related literature \cite{Bordini2006,treuil2008modelisation,michel2018multi}.
  \TinyTroupe{}'s principles seen in Section \ref{sec:principles}, particularly its programmatic style and ``toolkit-like'' organization, draws inspiration from some of these classical systems, though adapting them to language-centric problems.\footnote{As a side note, we've noticed that recent work on LLM-based agents hardly ever mentions this vast traditional pre-LLM MAS literature. Why that's the case is beyond our present scope, but we do encourage readers to familiarize themselves with this body of knowledge in case they are unaware of it. Current LLM technology probably offers fruitful ways to build on top of that, as \TinyTroupe{} itself shows.}

  The recent surge of LLMs has brought renewed interest and novel approaches in research and tooling \cite{10.24963/ijcai.2024/890}. Generative Agents \cite{generativeagents} is an influential LLM-based multiagent simulation approach that inspired many others \cite{qian2024chatdev,li2024agent,fan2024ai,shinn2023reflexion,chan2023chateval,packer2023memgpt,vezhnevets2025multi}, including the present work. It introduces the use of LLMs to produce agent behavior within an environment where they interact with one another in game-like scenarios. \TinyTroupe{} can be seen as an extension of this approach: it borrows key concepts, including the essential process of generating simulated agent behavior via LLM prompting; and it adds, on top of that, a layer of programmatic, analytical and experimentation concepts, aiming at business applications and other interactive scenarios of practical interest.

  Producing realistic human behavior remains a challenge. Park \emph{et al.} \cite{park2024generative} proposes an alternative to persona specification. They show how long, well-crafted and structured interviews can provide the basis for more realistic behavior by comparing the synthetic responses for standard psychological tests to the actual responses given by the corresponding interviewee. We hypothesize that such interviews could be transformed into (complex) persona specifications to be used with \TinyTroupe{}, but it is unclear to what extent. The environment itself can be challenging to simulate. \TinyTroupe{} uses \techterm{Intervention} and \techterm{TinyStory} to provide richer automated environmental inputs during the simulation. An alternative or complementary approach is to use a more general LLM-powered component to handle environmental interactions, like Concordia's \textit{Game Master} \cite{vezhnevets2025multi}.

Objectively evaluating such multiagent simulations in a general way is a challenge as well. While there are benchmarks for various other AI applications, there hardly seems to be anything equivalent for agent simulation libraries and frameworks, \textsc{PersonaGym}
\cite{samuel2024personagym} being a recent exception. Evaluating \TinyTroupe{} using this or similar methodologies, however, remains work to be done. Instead, as seen in Section \ref{sec:evaluation}, besides comparing simulations to actual human behavior, we also used an LLM-as-a-Judge technique, in general known to be fairly reliable \cite{zheng2023judging,gu2024survey,madaan2023self}, against several runs of our examples. The properties we measured and evaluated align with some of the common failure modes investigated by Cemri \emph{et al.}\cite{cemri2025multi}, notably \emph{disobey role specification} (e.g., persona adherence failure), \emph{disobey task specification} (e.g., converge discussions despite explicit instructions to diverge) and \emph{step repetition} (e.g., a surprisingly typical fluency problem), which suggests they are good quality markers to monitor. See also Arcuschin \emph{et al.}\cite{arcuschin2025chain}.

  Theoretical investigations about LLM-based multiagent systems have been accompanied by related tools, notably AutoGen \cite{wu2023autogen} and CrewAI \cite{crewai}. AutoGen is a library for problem-solving with multiagent systems in general, based on conversational patterns to constraint and guide the system's execution. CrewAI is similar, but its focus is on providing agents with \emph{background stories} and defining \emph{tasks} for them to solve (e.g., a programmer is more likely to handle programming tasks). \TinyTroupe{}, by contrast, is an LLM-powered technology fully dedicated to simulations, which, as we have seen, implies many unique requirements, mechanisms and applications, such as much longer and detailed persona specification to mirror the behavior of human customers. To the best of our knowledge, no other general open-source LLM-based multiagent simulation library currently provides equivalent capabilities, though there are specialized simulators, such as OASIS \cite{yang2024oasis} for large-scale social media simulation. Some new companies now commercialize products for similar aims, but their closed and proprietary nature makes any comparison difficult.\footnote{For instance: \url{aaru.com}, \url{syntheticusers.com}, \url{delphi.ai}, \url{askrally.com}, \url{atypica.ai} and \url{simile.ai}.}

\section{Conclusion}
\label{sec:conclusion}
 Our purpose in this work was to: show the unique requirements and possibilities of LLM-powered persona simulation, which markedly differ from problem-solving or assistive AI; to address these, introduce \TinyTroupe{}, an experimental simulation ``toolkit'', whose basic tenets, current elements and uses we have reviewed; and to demonstrate qualitatively its usefulness via concrete use cases. 
 We also performed selected empirical and quantitative evaluations, which revealed positive qualities, but also negative ones and unexpected trade-offs, among other phenomena. Deeper and more comprehensive analyses are necessary, but left as future work.
   
  As an ongoing project, many potentially interesting concepts and mechanisms remain to be added, for instance: improved or new memory types, such as the addition of \emph{procedural memories}; learning capabilities via Reinforcement Learning \cite{shinn2023reflexion}; more structured environment types, for instance hierarchical ones; corresponding factories to facilitate the setup of such environments; larger sets of standard \techterm{TinyTool}s; alternative or improved action correction mechanisms; integration with external systems, such as Web application control for testing purposes; and support for other LLMs beyond those offered by OpenAI.\footnote{Being an open-source project, the community is also welcome to propose such additions, as well as new use cases to challenge the current limits and motivate improvements. Conversely, we also hope that \TinyTroupe{} can provide inspiration and conceptual foundations for other projects.}

  Finally, as discussed in the beginning, so far we have not investigated the use of different LLMs. This, however, could be an effective way to produce more realistic simulation results. From our experiments, we know that the \texttt{GPT-5-mini} and other models of the \texttt{GPT-4/5} family are not always suitable for persona simulations, particularly when personas deviate from these models' biases (e.g., helpfulness, politeness), which is why we introduced an action correction mechanism. It is possible that other standard LLM models could help, though the fact that most of the main alternatives are also inherently designed for assistive AI makes this unlikely. More importantly, fine-tuning existing models to the persona simulation task could provide very significant gains at a relatively low cost, as suggested by Zhu \emph{et al.} \cite{zhu2024personality}. One way to produce the necessary data for this fine-tuning could be \TinyTroupe{} itself, by: simulating scenarios in which the action correction works well; manually manipulating specific trajectory points (e.g., force the agent to \texttt{THINK} some convenient thought via the \texttt{.think()} \techterm{TinyPerson} method); introducing branching simulation steering mechanisms (e.g., a modified \techterm{TinyStory}) in order to produce \emph{simulation trajectory trees} instead of mere linear trajectories; or even by using egregiously bad simulations as \emph{negative} training instances.

 \ifdoubleblind{}{
 
 \section*{Acknowledgments} 
 We would like to thank those who have helped or contributed to \TinyTroupe{} overtime, both directly and indirectly: Nilo Garcia Silveira, Olnei Fonseca, Bryant Key, Carlos Costa, Barbara da Silva, and other Microsoft colleagues.
 
 }

\bibliographystyle{plain}
\bibliography{tinytroupe}

\end{document}